\documentclass[twocolumn,citeautoscript,showpacs,preprintnumbers,amsmath,amssymb,prl,floatfix,footinbib]{revtex4}
\usepackage{graphicx}
\usepackage{dcolumn}
\usepackage{bm}
\usepackage{times}
\begin{document}
\title{Anomalous suppression of the orthorhombic distortion in superconducting Ba(Fe$_{1-x}$Co$_x$)$_2$As$_2$}
\author{S. Nandi, M. G. Kim, A. Kreyssig, R. M. Fernandes, D. K. Pratt, A. Thaler, N. Ni, S. L. Bud'ko, P. C. Canfield, J. Schmalian, R. J. McQueeney and A. I.
Goldman}\email{goldman@ameslab.gov}
\affiliation{\\Ames Laboratory, U.S. DOE and Department of Physics and Astronomy\\
Iowa State University, Ames, IA 50011, USA}

\date{\today}

\begin{abstract}
High-resolution x-ray diffraction measurements reveal an unusually
strong response of the lattice to superconductivity in
Ba(Fe$_{1-x}$Co$_x$)$_2$As$_2$. The orthorhombic distortion of the
lattice is suppressed and, for Co-doping near $x$~=~0.063, the
orthorhombic structure evolves smoothly back to a tetragonal
structure. We propose that the coupling between orthorhombicity and
superconductivity is indirect and arises due to the magneto-elastic
coupling, in the form of emergent nematic order, and the strong
competition between magnetism and superconductivity.
\end{abstract}


\maketitle The interplay between superconductivity, magnetism and
structure has become a major theme of research in the iron arsenide
families\cite{kamihara_08,rotter_08} of superconductors.  The strong
coupling between magnetism and structure, for example, is
illustrated by the parent compounds, AEFe$_2$As$_2$ (AE = Ba, Sr,
Ca), which manifest simultaneous transitions from a paramagnetic,
tetragonal phase to an antiferromagnetically ordered, orthorhombic
phase\cite{huan_08,jesche_08,goldman_08}. Strong coupling is also
evidenced by recent inelastic x-ray\cite{han_09} and
neutron\cite{mittal_09} scattering measurements of lattice
excitations, and Raman spectroscopy\cite{chauv_09}; all require
consideration of magnetic ordering or fluctuations to obtain
reasonable agreement with theory. Strong coupling between
superconductivity and magnetism are observed in several inelastic
neutron scattering
measurements\cite{chris_08,chi_09,lumsden_09,Li_09,inosov_09,pratt_09,chris_09_01},
which highlight the appearance of a resonance, or opening of a spin
gap, in the magnetic fluctuation spectrum below the superconducting
transition ($T_c$) in doped AEFe$_2$As$_2$ and LaFeAsO compounds.
Perhaps most striking, is the observation that the static magnetic
order for Co-doped BaFe$_2$As$_2$ is significantly suppressed below
$T_c$\cite{pratt_09,chris_09_01}.

Here we describe high-resolution x-ray diffraction measurements that
demonstrate an unusually strong response of the lattice to
superconductivity in Ba(Fe$_{1-x}$Co$_x$)$_2$As$_2$. Below $T_c$,
the orthorhombic distortion of the lattice is significantly
suppressed and, for $x~\approx~0.063$, the orthorhombic structure
evolves smoothly back to a tetragonal structure. Our observations
are consistent with a strong magnetoelastic coupling, realized
through emergent nematic order that manifests orientational order in
the absence of long-range positional order, analogous to the nematic
phase in liquid crystals. For the iron arsenide compounds, the
nematic phase corresponds to orientational order between two
antiferromagnetic sublattices, with staggered magnetizations,
\textbf{m}$_{1}$ and \textbf{m}$_{2}$, that are only weakly coupled
because of frustration that arises from large next-nearest-neighbor
magnetic interactions\cite{chandra_90,yildirim_08,si_08}. The
nematic order parameter is defined as
$\varphi=\textbf{m}_{1}\cdot\textbf{m}_{2}$. Above the structural
transition at $T_S$, the time-averaged quantities,
$\left<\varphi\right>=0$ and
$\left<\textbf{m}_{1}\right>=\left<\textbf{m}_{2}\right>=0$. With
the onset of nematic ordering at $T_S$, $\left<\varphi\right>\neq0$,
while $\left<\textbf{m}_{1}\right>=\left<\textbf{m}_{2}\right>=0$,
resulting in nematic order, but no static magnetic order.  Below
$T_N$, $\left<\varphi\right>\neq0$ \emph{and} static magnetic order
sets in with $\left<\textbf{m}_{1}\right>\neq0$ and
$\left<\textbf{m}_{2}\right>\neq0$. The nematic degree of freedom
leads to the structural distortion which lifts the magnetic
frustration\cite{fang_08,xu_08,fernandes_09}. Within this picture,
the competition between the orthorhombic distortion and
superconductivity is rooted in the coupling between magnetism and
superconductivity\cite{pratt_09,chris_09_01}, i.e. there is a common
origin for the suppression of both the structural and magnetic order
parameters below $T_c$.

\begin{figure}
\begin{center}
\includegraphics[clip, width=.45\textwidth]{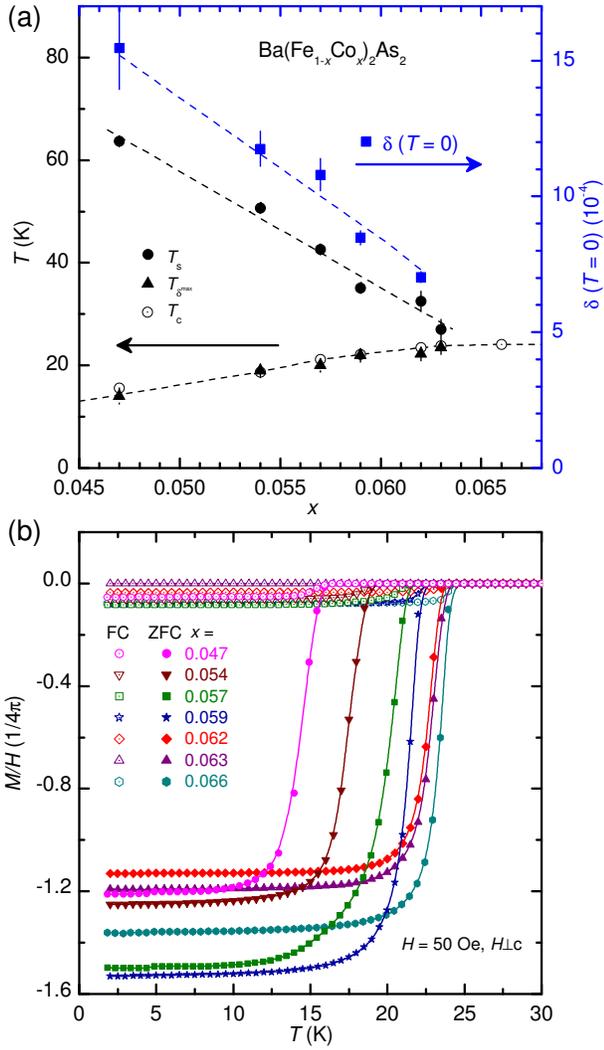}\\
\caption{(color online) (a) Tetragonal-to-orthorhombic transition
temperature, $T_S$, (filled circles) and superconducting transition
temperature, $T_c$, (open circles) from the present measurements as
a function of the average value of Co-doping as determined from WDS.
The filled triangles and squares represent $T_\delta$$^{max}$ and
$\delta$(0) as described in the text. Dashed lines serve as guides
to the eye. (b) The zero-field cooled and field cooled magnetization
for all samples in this study.} \label{magnetization}
\end{center}
\end{figure}

Single crystals of Ba(Fe$_{1-x}$Co$_x$)$_2$As$_2$ were grown out of
a FeAs self-flux using conventional high temperature solution
growth\cite{ni_08}. Details of the growth procedures are provided in
Ref.~\onlinecite{ni_08}.  The compositions were measured at between
10 and 30 positions on samples from each growth batch using
wavelength dispersive spectroscopy (WDS). The combined statistical
and systematic error on the Co composition is not greater than 5$\%$
(e.g. 0.063$\pm$0.003).  Since we have studied a set of samples with
Co concentrations within the error stated above, it is important to
establish that there are, indeed, systematic variations in sample
properties across our compositional range.  To this end, in
Fig.~\ref{magnetization}(a), we plot both the
tetragonal-to-orthorhombic transition temperature, $T_{S}$,
determined from our x-ray measurements and $T_c$, as determined from
the onset of the superconducting transition in the magnetization
data in Fig.~\ref{magnetization}(b). These data clearly establish
that both $T_c$ and $T_{S}$ change systematically with the average
Co composition determined by WDS, and we will employ these values in
the remainder of this paper.  We further note that the structural,
compositional, thermodynamic and transport measurements on samples
from each batch are consistent with the data presented in
Ref.~\onlinecite{ni_08}. Figure~\ref{magnetization}(b) shows that
all samples exhibit relatively sharp superconducting transitions and
we find no anomalies or measurable changes in the magnetization at
temperatures below $T_c$, indicating that the superconducting volume
fraction does not evolve below $T_c$.

Temperature-dependent, high-resolution, single-crystal x-ray
diffraction measurements were performed on a four-circle
diffractometer using Cu K$_{\alpha1}$ radiation from a rotating
anode x-ray source, selected by a germanium (1~1~1) monochromator.
For these measurements, the plate-like single crystals with typical
dimensions of 3\,$\times$\,3\,$\times$\,0.5\,mm$^3$ were attached to
a flat copper sample holder on the cold finger of a closed-cycle
displex refrigerator. The mosaicities of the
Ba(Fe$_{1-x}$Co$_x$)$_2$As$_2$ single crystals were all less than
0.02$^\circ$ full-width-at-half-maximum as measured by the rocking
curve of the (1~1~10) reflection at room temperature. The
diffraction data were obtained as a function of temperature between
room temperature and 8~K, the base temperature of the refrigerator.

\begin{figure}
\begin{center}
\includegraphics[clip, width=0.45\textwidth]{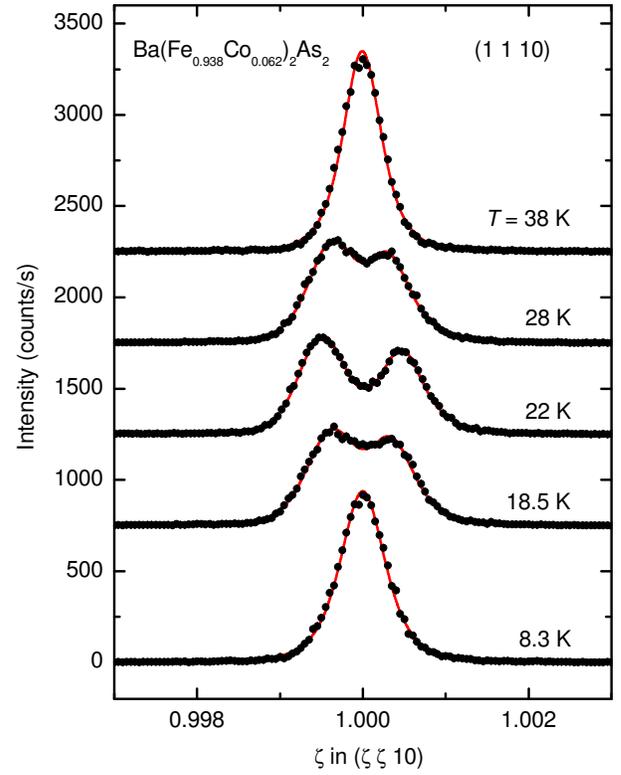}\\
\caption{(color online) Temperature evolution of the (1~1~10) Bragg
peak in Ba(Fe$_{0.938}$Co$_{0.062}$)$_2$As$_2$ at several
temperatures. The red lines represent fits to the data using either
one (for $T$~=~38~K and 8.3~K) or two (for $T$~=~28~K, 22~K and
18.5~K) Lorentzian squared peaks. For this sample, $T_S~=32$~K and
$T_c~=~23$~K.} \label{110}
\end{center}
\end{figure}

Figure~\ref{110} shows a subset of ($\zeta$~$\zeta$~0) scans through
the (1~1~10) reflection for Ba(Fe$_{0.938}$Co$_{0.062}$)$_2$As$_2$
as the sample was cooled through $T_{S}$~=~32$\pm$1~K.  The
splitting of the peak below $T_{S}$ is consistent with the
structural transition, from space group $I4/mmm$ to $Fmmm$, with a
distortion along the [1~1~0] direction. As the sample is cooled
further, the orthorhombic splitting increases until
$T_c$~=~23$\pm$1~K. Lowering the temperature below $T_c$ results in
a smooth decrease in the orthorhombic distortion until, below
approximately 13~K, a single component line shape reproduces the
data.


\begin{figure}
\begin{center}
\includegraphics[clip, width=0.45\textwidth]{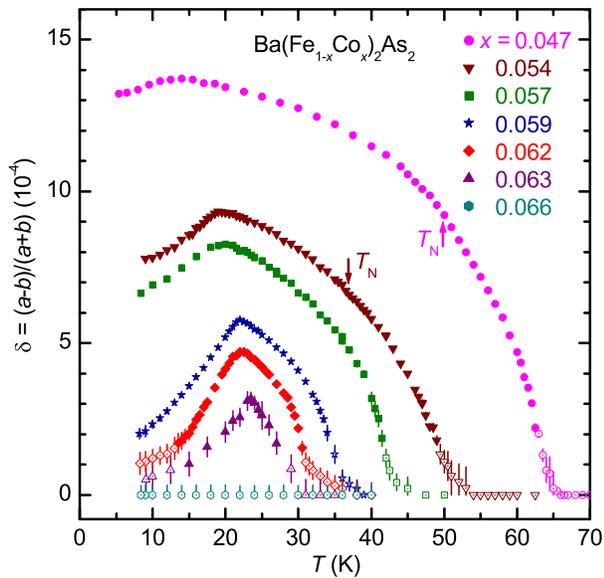}\\
\caption{(color online) The measured orthorhombic distortion
$\delta$ as a function of temperature. Filled symbols represent the
distortion determined from the positions of two peak fits to the
data. The open symbols represent an upper limit on the distortion
extracted from the line broadening of a single peak fit to the data
relative to the peak width well above $T_S$. Labeled arrows denote
the measured $T_N$ for several samples.} \label{ortho}
\end{center}
\end{figure}

In order to map systematic changes in structure with composition,
the temperature dependence of the orthorhombic distortion,
$\delta=\frac{(a-b)}{(a+b)}$, was measured for a series of seven
Ba(Fe$_{1-x}$Co$_x$)$_2$As$_2$ samples, with
0.047~$\leq$~$x$~$\leq$~0.066, and is displayed in Fig.~\ref{ortho}.
The solid symbols in Fig.~\ref{ortho} represent $\delta$, as
determined from fits to the ($\zeta$~$\zeta$~0) scans using two
Lorentzian squared peaks. For the $x$~=~0.062 and $x$~=~0.063
samples at low temperature, however, a single peak was sufficient,
as demonstrated in Fig.~\ref{110}. The open symbols in
Fig.~\ref{ortho} represent an upper limit on $\delta$ based on the
residual broadening of a single peak fit to the data, with respect
to the peak width determined from scans well above $T_{S}$.

The relative decrease in the orthorhombicity below $T_c$ is
pronounced and increases with increased doping.  Indeed, the
$x$~=~0.063 sample exhibits reentrant behavior, within experimental
uncertainty, where the low-temperature structure returns to
tetragonal symmetry below $T_c$. For $x$~=~0.066, no transition to
the orthorhombic structure was observed, defining an upper
Co-concentration limit for the tetragonal-to-orthorhombic phase
transition. To place the magnitude of the suppression of the
orthorhombicity in some context, we note that ultrahigh-resolution
thermal expansion measurements on untwinned single crystals of
YBa$_2$Cu$_{3}$O$_{7-\delta}$ also found a change in the
orthorhombic distortion at $T_c$, but smaller than the present case
by approximately two orders of magnitude\cite{Meingast_91}.

\begin{figure}
\begin{center}
\includegraphics[clip, width=0.45\textwidth]{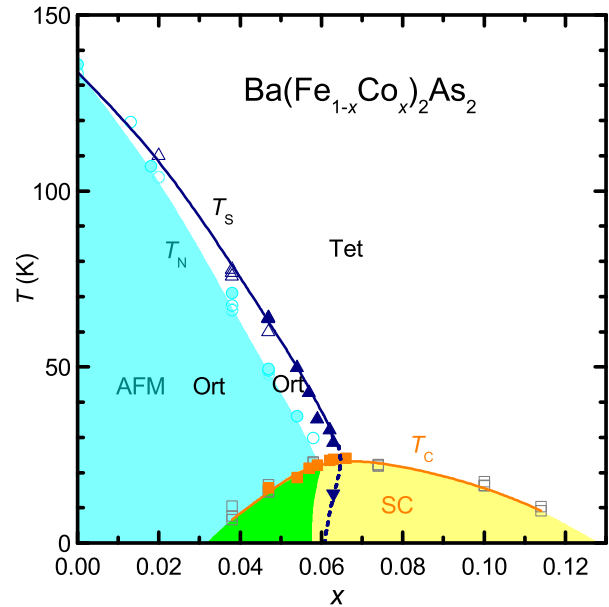}\\
\caption{(color online) The $\emph{T}$-$x$ phase diagram for
Ba(Fe$_{1-x}$Co$_x$)$_2$As$_2$ compiled from data in
Refs.~\onlinecite{ni_08} (open
symbols),~\onlinecite{fernandes_09_01} (filled symbols for $T_N$),
and the present study (filled symbols for $T_S$ and $T_c$). The
extension of the tetragonal-to-orthorhombic phase line into the
superconducting dome is represented by the dashed line.}
\label{Phase}
\end{center}
\end{figure}

With these results in hand, we have refined the phase diagram, shown
in Fig.~\ref{Phase}, to indicate how the phase line representing the
tetragonal-to-orthorhombic transition bends back below $T_c$. We
also plot, in Fig.~\ref{magnetization}(a), both the temperature,
$T_\delta$$^{max}$, at which the orthorhombic distortion for a given
sample is at a maximum, and $\delta$(0), the $T$~=~0 extrapolated
value for $\delta$ determined from a power law fit to the data above
$T_c$. We find that $T_\delta$$^{max}$ is coincident with $T_c$ for
all samples, and the monotonic decrease in $\delta$(0) with
increasing Co-doping is consistent with the decrease in $T_{S}$ for
each sample.  Furthermore, an extrapolation of the dashed line to
$x$~=~0 finds agreement with the value of $\delta$(0) for the parent
BaFe$_2$As$_2$ compound.

The strong suppression of the structural order parameter at low
temperatures is highly unusual and clearly connected to the onset of
superconductivity. The leading coupling in a Landau expansion of the
free energy between the orthorhombic distortion $\delta$ and the
superconducting order parameter $|\Psi|$ is
$\frac{\gamma_{\delta}}{2}\delta^{2}|\Psi|^{2}$. In principle, one
could rationalize our results as arising from a strong competition
between orthorhombic order and superconductivity. This would then be
reflected in a coupling constant,$\gamma_{\delta}$, sufficiently
large to suppress $\delta$ below $T_c$ but sufficiently small to
avoid a first order transition between both states. The temperature
variation of $\delta$ is, however, very reminiscent of the behavior
of the ordered magnetic moment, which has been shown to be strongly
suppressed below $T_c$ in Refs.~\onlinecite{pratt_09}
and~\onlinecite{chris_09_01}. Understanding both phenomena would
require the simultaneous fine tuning of the phase competitions, i.e.
of $\gamma_{\delta}$ and the corresponding coupling constant
$\gamma_{m}$, describing the interaction between magnetism and
superconductivity via
$\frac{\gamma_{m}}{2}(\textbf{m}_{1}^{2}~+~\textbf{m}_{2}^{2})|\Psi|^{2}$,
where $\textbf{m}_{1}$ and $\textbf{m}_{2}$ are the staggered
magnetizations corresponding to the two Fe-sites in the basal plane.

An intriguing alternative explanation of our results is rooted in
the unusual magneto-elastic coupling of the iron arsenides and the
competition between superconductivity and magnetism.  First, we
again note that commensurate antiferromagnetic fluctuations, of the
kind seen in the iron arsenides, have been shown to lead to an
emergent, nematic order parameter\cite{chandra_90},
$\varphi=\textbf{m}_{1}\cdot\textbf{m}_{2}$.  As discussed in detail
in Refs.~\onlinecite{fang_08,xu_08,fernandes_09}, $\textbf{m}_{1}$
and $\textbf{m}_{2}$ are weakly coupled and their relative
orientation is only fixed once nematic order sets in so that
$\left<\varphi\right>\neq0$. Nematic ordering occurs when the
correlation length for spin fluctuations, $\xi$, reaches a finite
threshold value\cite{fernandes_09}.  We emphasize here that nematic
ordering transpires via the spin fluctuations of the system and does
not require static magnetic ordering.  Therefore, nematic order, and
the associated structural distortion, discussed below, can occur
above the magnetic transition, or even in its absence.  This is
consistent with the fact that the tetragonal-to-orthorhombic
transition occurs at temperatures above the onset of static magnetic
order (see Fig.~\ref{Phase}) and our observation of a
tetragonal-to-orthorhombic transition for the $x$~=~0.063 sample in
the absence of static magnetic order down to
2~K\cite{fernandes_09_01}.

To understand the coupling between nematic order and the lattice
distortion we consider the leading terms in $\delta$ that contribute
to the free energy,
$\lambda\delta\varphi~+~\frac{C_{s,0}}{2}\delta^{2}$, with the bare
shear modulus, $C_{s,0}$, and coupling constant, $\lambda$.
Minimizing the free energy with respect to $\delta$ leads to the
relation $\delta~=~-\frac{\lambda}{C_{s,0}}\left<\varphi\right>$;
the lattice distortion simply follows the nematic order parameter.
Simultaneous orthorhombic and nematic ordering lifts the magnetic
frustration and allows magnetic long-range order.

Both the suppression of magnetic long range order below
$T_c$\cite{pratt_09,chris_09_01} and the suppression of the
orthorhombicity described here can be understood as a consequence of
the competition between itinerant magnetism and superconductivity
for the same electrons.  In the superconducting state, the magnetic
fluctuations are modified.  We have already noted, for example, the
opening of a gap and the appearance of a resonance in several
inelastic neutron scattering
measurements\cite{chris_08,chi_09,lumsden_09,Li_09,inosov_09,pratt_09,chris_09_01},
illustrating a change in the spin dynamics in the superconducting
phase. This competition between magnetism and superconductivity also
leads to a decrease in the spin fluctuation correlation length,
$\xi$\cite{fernandes_09}. From the discussion above, it then follows
that superconductivity weakens the magnetic spin fluctuations below
$T_c$, hence, suppressing the nematic order and the consequent
orthorhombic distortion.


In summary, our high-resolution x-ray diffraction measurements have
revealed an unusually strong response of the lattice to
superconductivity in Ba(Fe$_{1-x}$Co$_x$)$_2$As$_2$, leading to a
tetragonal structure for $x$~$\approx0.063$.  We propose that the
coupling between the orthorhombic distortion and superconductivity
is indirect and arises from the strong competition between magnetism
and superconductivity, together with a strong magnetoelastic
coupling in the form of emergent nematic order.  The appeal of this
scenario is that no new direct coupling between the elastic and
superconducting order parameters is required.
\\

We thank A. Kracher for the WDS measurements. The work at the Ames
Laboratory was supported by the US DOE, office of science, under
contract No. DE-AC02-07CH11358.

\bibliographystyle{apsrev}
\bibliography{Nandi_et_al_resub}

\end{document}